\documentstyle[preprint,aps]{revtex}
\begin{document}
\draft
\title{Do inertial electric charges radiate with respect to uniformly
accelerated observers?\footnote{This Essay was awarded a
{\em Honorable Mention} for 1994 by the {\em Gravity Research Foundation}}}
\author{George E. A. Matsas}
\address{Instituto de F\'\i sica Te\'orica,
         Universidade Estadual Paulista\\
         Rua Pamplona 145,
	 01405-900, S\~ao Paulo, SP\\
	 Brazil}

\maketitle
\begin{abstract}
We revisit the long standing problem
of analyzing an inertial
electric charge from the
point of view of uniformly accelerated observers in the context
of semi-classical gravity. We choose a
suitable set of accelerated observers with respect to which
there is no photon emission coming from the inertial charge.
We discuss this result against previous claims \cite{Roh}.
\end{abstract}
\narrowtext

This essay is devoted to call attention to the increasing
interplay between gravity and apparently unrelated areas.  It is
remarkable how mathematical techniques developed to shed light
on quantum gravity physics ended by improving and changing our
understanding about well established concepts. This is
particularly true for semi-classical gravity where many
subtleties about the elementary particle concept were realized.
In order to illustrate it, we shall discuss in the context of
semi-classical gravity the long standing problem  whether
inertial electric charges radiate or not with respect to
uniformly accelerated observers \cite{Roh}.

In the early sixties it was realized that radiation is not a
covariant concept. In 1963 Rohrlich found out that uniformly
accelerated electric charges should not emit radiation with
respect to co-accelerated observers
\cite{Roh}. This same conclusion was obtained lately by Boulware
who also discussed the problem using a classical approach
\cite{B}.  Recently, this issue was investigated
quantum-mechanically in the framework of semi-classical gravity
\cite{HMS1,HMS2}.  The result can be summarized in the
statement: {\em Every Minkowski photon emitted by an accelerated
charge as defined in the inertial frame corresponds to the
emission or absorption of a zero-energy Rindler photon as
defined in the accelerated frame.} This is in agreement with the
classical result because zero-energy Rindler photons are not
detectable by observers with finite acceleration \cite{HMS2}.
The {\em Minkowski} and {\em Rindler}  labels will be used to
distinguish between quantities defined with respect to inertial
and accelerated frames respectively.  This is necessary because
one of the main outputs obtained from semi-classical gravity is
that the particle content of a quantum field theory can be
distinct in diverse frames.

A related but different question is: {\em Do inertial electric
charges radiate with respect to uniformly accelerated
observers?} Rohrlich  addressed this question in the context of
classical electrodynamics \cite{Roh}.  Basically, he first
calculated the electromagnetic stress-energy tensor $T_M^{\mu
\nu}$ due to a static charge in the usual Minkowski coordinates
$(t,z,x,y)$, and next he transformed the result to Rindler
coordinates $(\tau, \xi, x, y)$, obtaining $T_R^{\mu \nu}$.
Rindler coordinates are the natural coordinates to study
uniformly accelerated observers, since they are static in this
coordinate system. Rindler and Minkowski coordinates are related
by
\begin{equation}
t = \frac{e^{a\xi}}{a}\sinh a\tau , \;
z = \frac{e^{a\xi}}{a}\cosh a\tau ,
\label{RC}
\end{equation}
where $a$ is some positive constant. Rohrlich states that the
presence of non-vanishing $T_R^{0 i}$ components indicates that
uniformly accelerated observers see radiation coming from
inertial charges. Yet mathematically correct, the interpretation
of this result is pretty anti-intuitive on energy grounds: {\em
Where does the radiating energy come from?}

By now, it is well known that the relevant manifold to quantize
any field with respect to uniformly accelerated observers is the
Rindler wedge \cite{F}, {\em i.e.,} the portion of Minkowski
space defined by $z > \vert t \vert$. The Rindler wedge can be
covered by Rindler coordinates (\ref{RC}).  The main difficulty
in analyzing the radiation emitted by an inertial charge in
terms of photons is the fact that, {\em in general,} the current
associated with such a charge [$j(x^\mu)= e
\delta(x)\delta(y)\delta(z)$] cannot be completely confined in
any Rindler wedge.  This is so because this current does not have a
compact support in the $t$-time. As a result, it is not clear
how to define the radiation associated with an inertial charge
in terms of photons in an arbitrary uniformly accelerated frame.
Here we shall show, however, that an adequate choice of
accelerated observers can give a definite answer to this
question.  Namely, this set of accelerated observers would
ascribe no particle emission coming from the inertial charge,
and consequently no radiation.

A world line given by $\xi ,x , y = {\mbox{\rm const}}$ is
characterized by having a constant proper acceleration $a
e^{-a\xi}$.  It was recently shown that an electric charge $e$
following such a world line can only emit zero-energy Rindler
photons in the accelerated frame.  Yet
possessing vanishing frequency, zero-energy Rindler photons
carry transverse momentum ${\bf k}_\bot \equiv (k_x, k_y)$.  The
emission rate of such photons was calculated in a gauge
invariant way \cite{HMS1,HMS2}:
\begin{equation}
P_{k_\bot} d{\bf k}^2_\bot =
\frac{e^2}{4\pi ^3 a e^{- a \xi} }
\vert K_1(k_{\bot } e^{a \xi} /a) \vert^2  d{\bf k}^2_\bot ,
\label{P}
\end{equation}
where $K_\nu (z)$ is the Bessel function of imaginary argument.
Now, let us take a sequence of world lines $W_n: [x=y=0, \xi
=\xi_n = {\mbox{\rm const}}$ $(n \in {\bf N}) ]$, such that
$\xi_{n+1} > \xi_n$ and $\xi_{n \to +\infty} \to +\infty$.  A
charge following a world line $W_n$ will have constant proper
acceleration $\sqrt{a_\mu a^\mu} = ae^{-a\xi_n}$.  In
particular, a charge following the world line $W_{n \to
+\infty}: [ x=y=0$, $\xi = {\mbox{\rm const}} \to +\infty ]$
will be inertial, since $a_\mu a^\mu =0$, and will be confined
inside the Rindler wedge by construction. As a consequence, the
corresponding accelerated observers associated with the Rindler
wedge will be able to answer properly if they expect to detect
or not any photon emission from this inetial charge. It is this inertial
charge which will be chosen to be analyzed in the uniformly
accelerated frame.  Taking the limit $\xi \to +\infty$ in
(\ref{P}), we conclude that such an inertial charge emits {\em
no} photons at all with respect to our accelerated observers.
Our result is not in contradiction with Rohrlich's conclusion
because both approaches are inequivalent.  Note that in the
procedure above we decided to move the charge with respect to
the Rindler wedge.  This is completely equivalent of keeping the
charge still, and moving the accelerated observers to the
opposite direction.

The fact that these accelerated observers detect no radiation is compatible
with a number of other facts.  Firstly, the emission or
absorption of a Rindler photon as defined in the accelerated
frame is unavoidably associated with the emission of a Minkowski
photon as defined in the inertial frame {\em provided the source
is contained in the Rindler wedge}
\cite{UW,KW}. Since
inertial charges do not emit Minkowski photons with respect to
inertial observers, they cannot emit Rindler photons with
respect to our accelerated observers either. Notice that this
reasoning depends crucially of the fact that the whole current
is confined inside the wedge.  Secondly, it was natural to
expect on energy grounds that inertial charges do not emit
photons with respect to these observers because there is in this
case no way to provide energy for the radiation.

Although our conclusion that an inertial charge following $W_{n \to +\infty}$
should not radiate with respect to our uniformly accelerated
observers could be qualitatively anticipated from the first
argument given above, we believe that this discussion
contributes to clarify this question, and dissipates any
misconceptions about it.  Inequivalent definitions of radiation
are allowable, provided they do not disagree concerning any real
events. This is clearly illustrated by the Unruh effect
\cite{U}, which predicts that the excitation of a detector
uniformly accelerated in Minkowski vacuum is associated with the
emission of a Minkowski particle, and absorption of a Rindler
particle according to inertial and co-accelerated observers
respectively. Notwithstanding, both set of observers agree about
the excitation rate of the detector. It should be so because the
excitation phenomenon is an event, in opposition to the particle
content of a field theory which depends on the reference frame.
In the case studied above, every approaches must agree with the
fact that an inertial electric charge must stay at rest with
respect to, say, a companion uncharged particle. In our analysis
in the accelerated frame, this conclusion is straightforward
since, according to us, the inertial electric charge emits no Rindler
photons, and thus, no recoiling can be observed.

It is widely believed that semi-classical gravity can predict
quantum gravity effects before a full theory is available. It
may be that the Hawking radiation emitted by black holes turns out
to be the first observed quantum gravity effect.  Yet only
future data will be able to decide on it, it is fair to say that
semi-classical gravity has broadened considerably
our knowledge about many physical concepts, and is valuable
in its own right.

\begin{flushleft}
{\bf{\large Acknowledgements}}
\end{flushleft}

I am deeply indebted to Atsushi Higuchi and Daniel Sudarsky
for enlightening conversations about various topics related with
this issue, and to the inspiring enviroment provided by the
Enrico Fermi Institute at the University of Chicago where these
discussions began.  I am also thankful to Conselho Nacional de
Desenvolvimento Cient\'\i fico e Tecnol\'ogico, a Brazilian
agency, for financial support during the period this work was
produced.

\end{document}